\documentclass[twocolumn,aps,prb,superscriptaddress,amssymb,showpacs]{revtex4-1}
\usepackage[dvips,final]{graphicx}
\begin{document}

\title{Ballistic transport at room temperature in micrometer size multigraphene}

\author {S. Dusari}
\author{J. Barzola-Quiquia}
\author{P. Esquinazi}\email{esquin@physik.uni-leipzig.de}
\affiliation{Division of Superconductivity and Magnetism, Institut
f\"ur Experimentelle Physik II, Universit\"{a}t Leipzig,
Linn\'{e}stra{\ss}e 5, D-04103 Leipzig, Germany}

\author{N. Garc{\'i}a}
\affiliation{Laboratorio de F\'isica de Sistemas Peque\~nos y
Nanotecnolog\'ia,
 Consejo Superior de Investigaciones Cient\'ificas, E-28006 Madrid, Spain}


\begin{abstract}
The intrinsic values of the carriers mobility and density of the
graphene layers inside graphite, the well known structure built on
these layers in the Bernal stacking configuration, are not well
known mainly because most of the research was done in rather bulk
samples where lattice defects hide their intrinsic values. By
measuring the electrical resistance through microfabricated
constrictions in micrometer small graphite flakes of a few tens of
nanometers thickness we studied the ballistic behavior of the
carriers. We found that the carriers' mean free path is micrometer
large with a mobility $\mu \simeq 6 \times 10^6~$cm$^2$/Vs and a
carrier density $n \simeq 7 \times 10^8~$cm$^{-2}$ per graphene
layer at room temperature. These distinctive transport and
ballistic properties have important implications for understanding
the values obtained in single graphene and in graphite as well as
for implementing this last in nanoelectronic devices.
\end{abstract}

\maketitle

  The
existence of massless-like quasiparticles (Dirac fermions) found
in the quasi-two dimensional graphite structure
\cite{luky04,li07,orl08} and in single graphene layers
\cite{novo05,zhang05} enhanced enormously the interest in these
materials, implying not only the start of a qualitative new
physics\cite{cas09} but also the possibility of nanotechnological
improvements \cite{gei09,ouy10}. The search for highest carrier
mobility together with lowest carrier density in graphene remains
a main research issue of this two-dimensional model system also
because of their implications in future applications in
nanoelectronics.  It is nowadays known, however, that graphene
properties can be sensitively affected by the environment,
particularly by the substrate \cite{sab08,gar10} and by disorder
\cite{mor08}. This fact is supported by experiments done on
suspended graphene layers where mobility approaching $\sim 5
\times 10^5~$cm$^2$/Vs at carrier densities $n \sim
10^9~$cm$^{-2}$ at $T < 30~$K has been reached \cite{du08,bol08}.
Nevertheless, the expectations on the use of single graphene layer
(SGL) on substrates are high \cite{gei09} and in spite of the
detrimental substrate influence, SGL exhibits ballistic transport
in the 100~nm range at room temperature \cite{lin09} and in the
$\sim 1~\mu$m range at very low temperatures \cite{mia07}.

By measuring the resistance and its ballistic characteristics we
studied the transport behavior of the carriers of the graphene
layers inside the graphite structure. This study allows us to
obtain the carriers' values for the mean free path, carrier
density and mobility. Due to the natural packing of the graphene
layers inside graphite we expect not only a natural shielding from
the environmental influence but their should be actually more
ideal than SGL standing alone or on a substrate. Although some
experiments have been done in graphene bilayers, the actual
intrinsic values of these transport parameters  for the graphite
structure remain still not well known simply because their are
very sensitive to lattice defects \cite{arn09}, grain boundaries
\cite{gon07,gar08} and internal interfaces \cite{bar08}, which in
general trigger high carrier density with low mobility. Recent
results obtained on bulk oriented pyrolytic graphite samples
indicate ballistic transport in the micrometer range at $T < 10~$K
with a very low carrier density $n < 10^9~$cm$^{-2}$ and huge
mobility $\mu > 10^6~$cm$^2$/Vs \cite{gar08}.

The graphite flakes we report here were obtained by exfoliation of
a highly oriented pyrolytic graphite sample of ZYA grade. Using an
ultrasonic technique  we obtained several flakes that we selected
by measuring the resistance, its temperature dependence and
micro-Raman signals and taking into account the overall shape and
dimension of the sample, its surface and thickness. The Au
contacts for longitudinal resistance measurements were prepared
using electron beam lithography, see Fig.~1.
\begin{figure}[]
\begin{center}
\includegraphics[width=1\columnwidth]{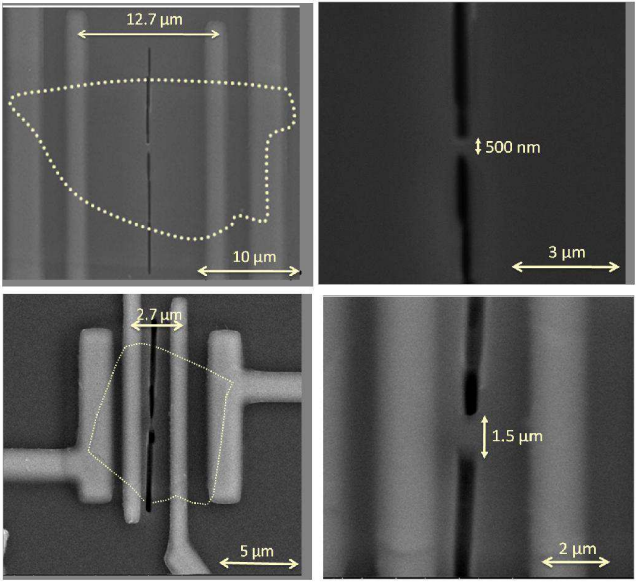}
\caption[]{Scanning electron microscope images of the two samples
discussed here with the four electrodes and with one of the
constrictions in between. The graphite flakes (white lines
indicate their perimeter) cannot be seen directly because of the
top 300~nm resin layer used to shield them from the Ga$^+$ focused
ion beam. Two upper pictures correspond to sample A and bottom to
sample B.}
\end{center}
\end{figure}

The measurement of the carrier mean free path $\ell$ is not
straightforward and  in general the value obtained depends on
several others, not well known parameters within the selected
transport model, usually based on a Drude-Boltzmann approach.
However, in case of ballistic conduction there is a transparent
method to obtain $\ell$ without adjustable parameters,
successfully used for macroscopic graphite samples as described in
\cite{gar08}. The method is based on the measurement of the
longitudinal resistance as a function of the geometry of a
constriction located between the voltage electrodes (see Fig. 1).
We prepared these constrictions  with the focused ion beam of a
dual beam microscope. We avoided the modification of the
crystalline structure of the samples due to the ion beam spread by
protecting them  with a negative electron beam resist (AR-N 7500)
of $\sim 300$~nm thickness, a method successfully tested in
graphite and described in \cite{barnano10}.

As shown in \cite{gar08} the resistance of a graphite sample of
width $\Omega$, thickness $t$,  with a constriction of size $W$
and length $L$ connecting two half-parts of resistivity  $\rho$ is
given by:
\begin{equation}
R(T) =a\frac{\pi\rho(T)}{4Wt}\ell(T) + a \left .
\frac{2\rho(T)\gamma(\kappa)\ln(\Omega/W)}{\pi t} \right |_{W <
\Omega} + \frac{\rho(T)L}{W t}\,.
\end{equation}
At the rhs of Eq.(1) the first  term corresponds to the ballistic
Knudsen-Sharvin resistance \cite{gar08}; the second, logarithmic
term to the ohmic, spreading resistance in two-dimensions with the
smooth function $\gamma(\kappa=W/\ell) \simeq 1 -
0.33/\cosh(0.1\kappa) = 0.67\ldots 1$ for $\kappa = 0 \ldots
\infty$ \cite{wex66}. The last term is due to the ohmic resistance
of the constriction tube itself. The constant $a$ takes care of
the influence of the sample shape,  the topology  and the location
of the electrodes in the sample.  It is equal to 1 for the
geometry used in sample~A where the voltage electrodes are
deposited through the whole sample width and their distance is
$\sim \Omega \gg W$. This can be verified from the measurement of
the resistance as a function of $W$, as it is shown below. To keep
the third term, the ohmic contribution of the constriction, small
enough we prepared constrictions with length $L < 0.4~\mu$m. The
results presented here correspond to two samples, sample A and B
with different geometry and resistivity. Sample A (B) had a size
(distance between voltage electrodes $\times$ width $\times$
thickness) of $12.7 \times 16 \times 0.015~\mu$m$^3$ ($ 2.7 \times
9.2 \times 0.040~\mu$m$^3$) and a resistivity $\rho(300~$K)$=
89~\mu\Omega$cm ($18~\mu\Omega$cm), see Fig.~1.

\begin{figure}[]
\begin{center}
\includegraphics[width=1\columnwidth]{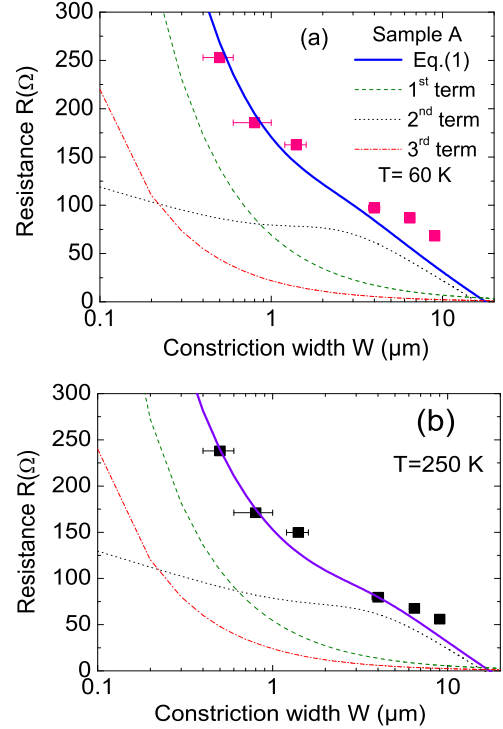}
\caption[]{Measured resistance for sample~A as a function of the
constriction width $W$ at (a) 60~K and (b) 250~K. The point with
the largest $W$ corresponds to the virgin sample without a
constriction. The different lines correspond to the first
ballistic term (dashed line) and the ohmic (2$^{\rm nd}$ (dotted
line) and 3$^{\rm rd}$ (dashed-dotted line) contributions in
Eq.~(1)) and the continuous line to the addition of the three. The
only free parameter is the mean free path $\ell$, see Eq.~(1). For
(a) the continuous line is calculated with $\ell = 1.2~\mu$m and
for (b) $\ell = 0.8~\mu$m.}
\end{center}
\end{figure}

The simplest and direct way to check the ballistic contribution
and obtain the mean free path (without further adjustable
parameter) is just measuring the resistance $R$ as a function of
the constriction width in combination with Eq.~(1). The results
for sample A at two temperatures are shown in Fig.~2. In that
figure one recognizes that for $W \leq 2~\mu$m the ballistic
contribution overwhelms the ohmic ones, see Eq.(1), indicating
that the mean free path should be of this order. The theoretical
lines in Fig.~2 were obtained using $\ell = 1.2~\mu$m and
$0.8~\mu$m at 60~K and 250~K, respectively. We estimate an error
of $\sim 35~\%$ for these values determined by the errors in the
measurement of the sample and constriction geometry. We can
calculate now the Fermi wavelength per graphene layer using the
relation
\begin{equation}
\lambda_F = 2 \pi e^2 N_s \rho \ell / h t\,
\end{equation}
with $N_s$ the number of graphene layers in the sample. For sample
A we obtain then $\lambda_F  = 0.5~(0.8)\pm 0.25~\mu$m at 250~K
(60~K).

In case $\lambda_F$ is larger than the constriction size, the
ballistic contribution to the resistivity is better described by
the inverse of a sum of an energy $E$- and transverse wave vectors
$q_n$-dependent transmission probabilities $T_n$, where $n = 0,
\pm 1, \pm 2, \ldots$ \cite{sta07}. The maximum possible $n$ is
determined by the constriction width $W$, decreasing the smaller
$W$. In this case the increase in resistance is expected to show
an oscillatory behavior as a function of $W$  or $\lambda_F$
\cite{gar89,sny08} as observed experimentally in Bismuth (Bi)
nanowires \cite{cos97} as well as in GaAs devices
\cite{wee88,wha88}.

Note that the distance between voltage electrodes in sample~A is
larger than the obtained $\ell$.  The larger the sample the larger
is the probability to have defective regions with larger carrier
concentration and smaller mean free path within the voltage
electrodes \cite{arn09}. Therefore we repeated the experiment with
sample~B that shows lower resistivity and with a smaller voltage
electrode distance, see Fig.~1. Figure~3 shows the measured
resistance normalized by its value at a constriction $W = 3~\mu$m;
in this way we pay attention to the huge relative increase
decreasing $W$ and we need neither the absolute value of $\rho$
nor of $a$ to compare the data with theory.

\begin{figure}[]
\begin{center}
\includegraphics[width=1\columnwidth]{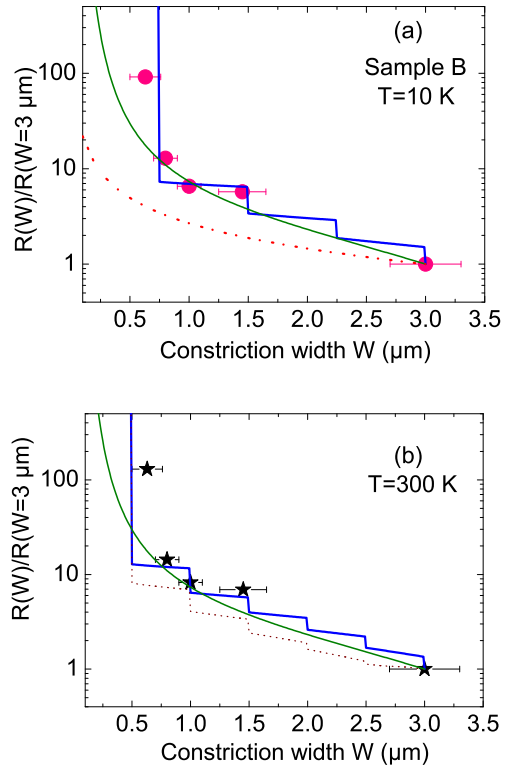}
\caption[]{Normalized resistance for sample~B vs. constriction
width $W$; note the semilogarithmic scale. (a) Data at 10~K. The
line with steps is obtained dividing the ballistic term in Eq.~(1)
by $(\lambda_F/2W)$trunc$(2W/\lambda_F)$ with parameters $\ell =
2.7~\mu$m and $\lambda_F = 1.5~\mu$m. The continuous line is
obtained multiplying the ballistic term in Eq.~(1) by the
exponential function $\exp(\lambda_F/2W)$. The dotted line follows
Eq.~(1) with $\ell = 2.7~\mu$m. (b) The same as in (a) but at
300~K. The used parameters are $\lambda_F = 1.0~\mu$m and $\ell =
2.2~\mu$m. The dotted stepped function is obtained using the same
$\lambda_F$ but with a smaller $\ell = 1.3~\mu$m.}
\end{center}
\end{figure}

\begin{figure}[]
\begin{center}
\includegraphics[width=1\columnwidth]{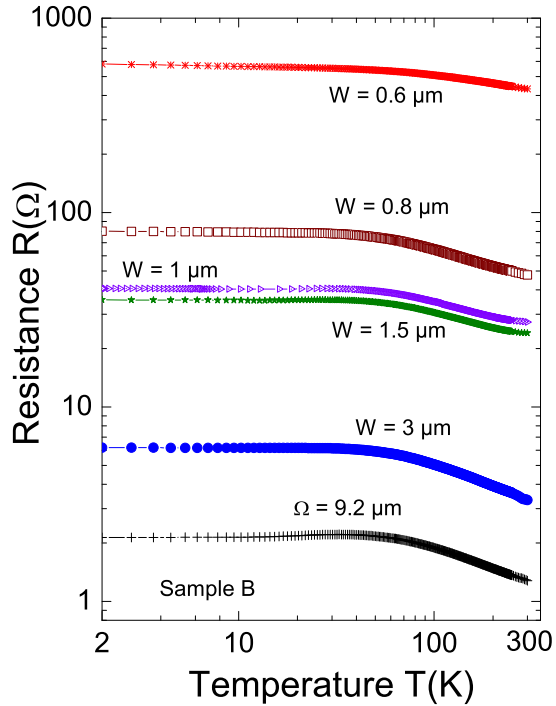}
\caption[]{Resistance of sample~B vs. temperature at different
constriction widths and without constriction ($(+), \Omega$ is the
total sample width, i.e. data of the sample without
constriction).}
\end{center}
\end{figure}

In Fig.~3 one realizes that for sample~B Eq.~(1) does not describe
the experimental data even assuming the largest possible mean free
path equal to the voltage electrode distance of $2.7~\mu$m. The
data can be reasonably well fitted dividing the ballistic term in
Eq.~(1) by the function ${\rm trunc}(2W/\lambda_F) \lambda_F/2W$,
which generates steps whenever the constriction width $W \simeq i
\lambda/2$ with $i$ an integer. From the fits we obtain the
parameters $\lambda_F = 1.0~ (1.5)~\mu$m and $\ell = 2.2~(2.7) \pm
0.3~\mu$m at 300~(10)~K, see Fig.~3. Using other values of $\ell$,
for example $\ell = 1.3~\mu$m, see Fig.~3, the function does not
fit the data indicating indeed that the carriers behave
ballistically between the voltage electrodes, leaving actually
$\lambda_F$ the only free parameter.

The ballistic analytical function we use resembles the theoretical
results with similar steps obtained in \cite{sny08} where the
conductance vs. $W$ was calculated numerically for a SLG  with an
electrostatically potential landscape that resembles a
constriction. An analytical average value or envelope of this
stepped  function is obtained replacing the truncation function by
$\exp(-\lambda_F/2W)$, see Fig.~3. This exponential function
represents the impossibility of an electron to propagate in the
constriction when $W < \lambda_F/2$, as also occurs for the
propagation of light in a tube.

The data appear to be  better fitted by the stepped function than
with the exponential one. The important result obtained for
sample~B is the huge increase of the resistance for $W < 2~\mu$m
indicating clearly a larger $\ell$. The larger $\ell$ in sample~B
to the one obtained in sample~A is compatible with the measured
resistivity difference.

As a further proof that the huge increase of the resistance
decreasing the constriction width in sample~B is due to the
ballistic contribution and not due to, e.g. a possible disorder
produced by the ion beam on the graphite structure, we show in
Fig.~4 the temperature dependence of the measured resistance
without and with the different constriction widths. We observe
that in spite of the huge resistance increase  the temperature
dependence remains similar. The small differences in the
temperature dependence of the resistance, decreasing slightly more
at $T > 80~$K at large values of $W$ than at the smallest $W$,
e.g. $R(2)/R(300) \simeq 1.6$ (sample without constrictions) to
$\simeq 1.45$ for $W = 0.6~\mu$m (see Fig.~4) can be explained
taking into account that the ballistic contribution (first term in
the r.h.s of Eq.(1)) is weakly $T$ dependent. At low values of $W$
the ohmic contributions start to be negligible compared with the
ballistic one (see Fig.~2) and therefore the temperature
dependence of $R(T)$ slightly reduces decreasing $W$, see Fig.~4.

As shown in \cite{gar08} the temperature dependence of $R(T,W)$
can be used now to obtain $\lambda_F(T)$ and the mobility per
graphene layer, this last given by $\mu(T) = (e/h) \lambda_F(T)
\ell(T)$. Since the density of carriers per graphene layer can be
calculated from $n = 2\pi/\lambda_F^2$ we show in Fig.~5 the
mobility vs. carrier density for the two samples and compare them
with data from literature for suspended SLG. From these results we
clearly recognize the much larger mobility and smaller density of
carrier for the thin graphite flakes, supporting the view that the
graphene layers within graphite are of better quality than SLG and
with a smaller carrier density. Obviously sample~B is  not free
from defects and therefore we expect that the obtained values
might still be improved in ideal, defect free graphite structures.

\begin{figure}[]
\begin{center}
\includegraphics[width=1\columnwidth]{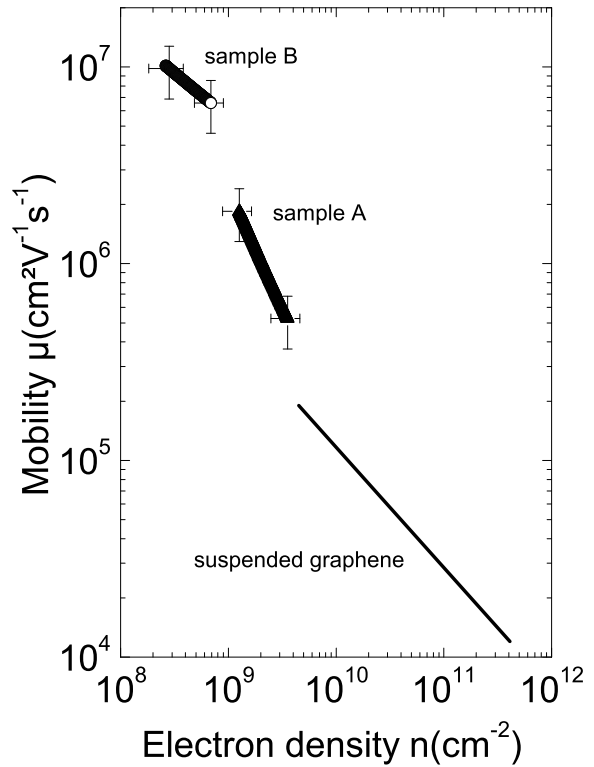}
\caption[]{Mobility vs. electron density obtained for samples A
and B. The points are obtained  between 300~K and 10~K (60~K for
sample A). The line corresponds to the data of a suspended SLG at
20~K from Ref.~\protect\onlinecite{du08}.}
\end{center}
\end{figure}

The rather weak temperature dependence of $n(T)$ (or $\lambda_F$)
obtained for the here reported samples is a non-trivial result
that deserves a comment. For the simple case of a linear
dispersion relation we expect that the $\lambda_F(T) \propto 1/T$
and therefore $n(T) \propto T^2$ or $\propto T$ for a quadratic
dispersion relation. The data obtained indicate that the carrier
density increases at most a factor of ten between 10~K and 300~K.
We may speculate the reason for this behavior, either by comparing
our results with those found in Bi and a possible band splitting
or the existence of an intrinsic energy gap in graphite.

Firstly, we note that both materials, graphite and Bi, have very
small carrier densities with  small anisotropic effective carrier
mass. For a long time it was unclear why there was such a
difference in the carriers' mean free path between the two
structures. While the carriers in Bi show $\ell (300~$K$)
> 2~\mu$m,\cite{bro70,yan00} the values reported for graphite in
literature are more than  one order of magnitude smaller at the
same temperature. The results of this work finally put an end to
this apparent difference showing that also in graphite the
carriers' mean free path can be micrometer large with a low
carrier density.

The large $\lambda_F$ of the order of the sample dimensions may
affect the band structure. For $\lambda_F \sim 1~\mu$m, and for
the typical sample dimensions we report here
 only the first five quantized wave vectors in reciprocal space
 are occupied ($E_F \sim 4~$meV) indicating that the usual
 continuous approximation of the band structure is inappropriate.
As in Bi nanowires we expect that quantum confinement affects the
electronic band structure and a split into subbands can occur
\cite{zha00}. It is interesting to note that the temperature
dependence of the resistivity of our samples, see Fig. 4, is
basically the same as for Bi nanowires of diameters $\phi < 70~$nm
\cite{zha00}. Therefore, it may be possible that the carriers in
the micrometer small graphite samples fill partially  a narrow
band with energy gap of the order of $\sim 10~$meV, preventing a
simple thermally activated excitation of the carriers to the
conduction band.

Because macroscopic graphite samples have a much larger density
defects ($n(T \gg 1~$K$) \gtrsim 10^{11}~$cm$^{-^2}$ per graphene
layer) than our samples \cite{gar08}, it appears plausible that
the temperature behavior of the resistance observed here cannot be
seen in bulk HOPG. Experimental evidence for the influence of
interfaces to the transport properties of HOPG was reported
recently  \cite{bar08} supporting this statement. It remains
unclear what we expect to happen in ideal graphene. The reported
data in literature were taken mostly for samples with larger
carrier density than in our samples. Therefore, we believe that
one cannot clearly answer yet on the existence or not of an energy
gap in ideal graphene (experiments at carrier concentration $<
10^8$~cm$^{-2}$ were not yet reported). At such low carrier
densities, as it appears to manifest in graphite, electron
correlations and possible localization effects should be
considered. Electron interactions are large and for a small enough
carrier density, the expected screening will be very weak
promoting therefore the existence of an energy gap. This is what
it is observed in Monte Carlo simulations for the unscreened
Coulomb interaction in graphene with Dirac flavor $N_s=2$, see
Ref.~\onlinecite{dru09}.

The obtained mobility in our thin micrometer small graphite
samples is two orders of magnitude larger,  with a carrier density
three orders of magnitude smaller  than those measured in bulk
samples, see e.g. Ref.~\onlinecite{sch09}, indicating that
macroscopic samples do not provide necessarily the intrinsic
properties of graphite due in part to the large density of lattice
defects \cite{arn09}. The ballistic behavior obtained here with
micrometer large mean free path and low carrier density  was
further verified by magnetoresistance and Hall effect measurements
and by the observation of Aharonov-Bohm oscillations in the
magnetoresistance.

The magnetoresistance for fields normal to the graphene planes
systematically decreases decreasing the constriction width,
starting already with the largest $W$ as shown in Fig.~6 for
sample~A (similar behavior is obtained for sample~B), as observed
in bulk HOPG samples but at lower temperatures \cite{gon07}. This
behavior indicates that $\ell \gtrsim 1~\mu$m up to room
temperature. The field dependence of the Hall resistance as well
as its temperature dependence in similar graphite flakes as
presented in this work, deviate clearly from measurements in bulk
graphite samples, see e.g. Ref.~\onlinecite{heb05}. Although the
calculation of the carrier density through the Hall effect in
semimetals due to the electron and hole contributions is non
simple, the obtained results support not only the existence of the
low carrier density but also the weak temperature dependence of
$n$. Finally, for fields parallel to the graphene planes and input
current the magnetoresistance shows pronounced Aharonov-Bohm
oscillations, similar to that found in Bi nanowires \cite{nik08}
and topological insulators \cite{pen09}, demonstrating the
coherent propagation of electrons around a length of several
micrometers up to temperatures $T \sim 250$~K.\cite{schade}

\begin{figure}[]
\begin{center}
\includegraphics[width=1\columnwidth]{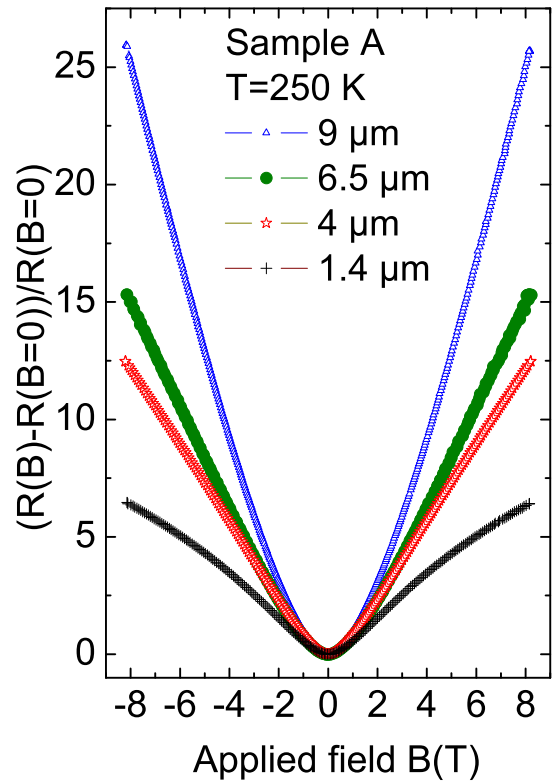}
\caption[]{Magnetoresistance vs. magnetic field applied normal to
the main surface of sample~A for different constriction widths and
at constant temperature of 250~K. The clear reduction of the
magnetoresistance decreasing $W$ supports the existence of a mean
free path of the order of several microns.}
\end{center}
\end{figure}


We thank  H. Peredo for technical support in the Hall effect
measurements. This work is supported by the Deutsche
Forschungsgemeinschaft under contract DFG ES 86/16-1; S.D. is
supported by the Graduate School  of Natural Sciences ``BuildMoNa"
of the University of Leipzig.


\end{document}